\documentclass[twocolumn ]{aa}
\usepackage{graphicx}
\usepackage{natbib}
\bibpunct{(}{)}{;}{a}{}{,}
\usepackage{graphicx}
\usepackage[mathcal]{eucal}
\usepackage{txfonts}
\begin{document}
   \title{A closure model with plumes}

   \subtitle{I. The solar convection}
    \author{K.~Belkacem\inst{1} 
            \and
            R.~Samadi\inst{1} 
            \and
            M.~J. Goupil\inst{1} 
            \and 
            F.~Kupka\inst{2} }

 \institute{
Observatoire de Paris, LESIA, CNRS UMR 8109, 92195 Meudon, France \and
Max-Planck-Institute for Astrophysics, Karl-Schwarzschild Str. 1, 85741 Garching, Germany }
\offprints{K. Belkacem}
\mail{Kevin.Belkacem@obspm.fr}
\date{
 Received 06 April 2006 / Accepted 18 August 2006}

\titlerunning{A closure model with plumes}

\abstract
{ Oscillations of stellar $p$~modes, excited by turbulent convection, are investigated.
In the uppermost part of the solar convection zone, radiative cooling is
responsible for the formation of turbulent plumes, hence the medium is modelled
with downdrafts and updrafts.}
{We take into account the asymmetry of the up- and downflows
created by turbulent plumes through an adapted closure model. In a companion
paper, we apply it to the formalism of excitation of solar $p$~modes developed
by Samadi \& Goupil (2001).}
{Using results from 3D numerical simulations of the uppermost part of the solar
convection zone, we show that the two-scale mass-flux model (TFM) is valid only
for quasi-laminar or highly skewed flows (Gryanik \& Hartmann 2002) and does not
reproduce turbulent properties of the medium such as velocity-correlation
products. We build a generalized two-scale mass-flux Model (GTFM) model
that takes both the skew introduced by the presence of two flows
\emph{and} the effects of turbulence in each flow into account. In order to apply the
GTFM to the solar case, we introduce the plume dynamics as modelled by
Rieutord \& Zahn (1995) and construct a closure model with plumes (CMP).}
{ The CMP enables expressing the third- and fourth-order correlation products in
terms of second-order ones. When compared with 3D simulation results, the CMP
improves the agreement for the fourth-order moments by a factor of 
two approximately compared with the use of the quasi-normal approximation 
or a skewness computed with the classical TFM.}
{ The asymmetry of turbulent convection in the solar case has an
important impact on the vertical-velocity fourth-order moment, which has to be
accounted for by models. The CMP is a significant improvement
and is expected to improve the modelling of solar $p$-mode excitation.

   }

   \maketitle
%

%
\section{Introduction}
In the uppermost part of the solar convective zone, turbulent entropy
fluctuations and  motions of eddies drive acoustic oscillations. 3D numerical
simulations of the stellar turbulent outer layers have been used to compute
the excitation rates of solar-like oscillation modes \cite{Stein01A}. As an
alternative approach, semi-analytical modelling can provide an understanding
of the physical processes involved in the excitation of $p$~modes: in this case,
it is indeed rather easy to isolate the different physical mechanisms at work
in the excitation process and  to assess their effects. 
Various semi-analytical approaches  have been developed by several authors
\citep{GK77,GK94,B92,Samadi00I}; they differ from each other by the nature of
the assumed  excitation sources, by the assumed simplifications and approximations, 
and also by the way the turbulent convection is described
(see the review by \citealt{Stein04}). 
Among the different theoretical approaches, that of \cite{Samadi00I} includes
a detailed treatment of turbulent convection, which enables us to investigate
 different assumptions about turbulent convection in the outer layers of stars
\citep{Samadi05c}. In this approach, the analytical expression for the acoustic
power supplied to the $p$~modes involves fourth-order correlation
functions of the turbulent Reynolds stress and the entropy source term, which
for the sake of simplicity are expressed in terms of second-order moments by
means of a closure model.

The most commonly used closure model at the level of fourth-order moments (FOM)
is the {\it Quasi-Normal Approximation} (QNA), which is valid for a Gaussian
probability distribution function \citep[see][]{Lesieur97} and was first
introduced by \cite{Million41}. The QNA is rather simple and convenient to
implement. However, \cite{Ogura63} has shown that such a closure could lead
to  part of the kinetic energy spectrum becoming negative. In this paper, we confirm
 the results of \cite{KR2006} (hereafter KR2006), namely that this approximation
indeed provides a poor description of the physical processes involved in
solar turbulent convection.

 Mass flux models (e.g., \citealt{Randall92}, \citealt{Ab}) explicitly
take the effects of {\it updrafts} and {\it downdrafts} on the
correlation products into account. The presence of two well-defined flow directions then introduces an additional 
contribution when averaging the fluctuating quantities, since averages of fluctuating
quantities over each individual flow differ from averages over the total flow. 
For applications in atmospheric sciences, the mass-flux model for convection
has recently been improved by \citet[][hereafter GH2002]{GH2002}. Their
motivation has been to account for the fact that horizontal scales of
temperature and velocity fluctuations are different (hence their improvements 
lead to a `two-scale mass-flux model' (TFM)) as well as to understand and measure the
effects of the skewness of their distribution. According to GH2002, mass-flux models, which also
include the TFM, underestimate the FOM by as much
as 70\%. Therefore, such models clearly miss some important physical
effects present in convective flows. \cite{GH2002} and \cite{GH2005} studied
the asymptotic limits of TFM which led the
authors to propose an interpolation between the QNA and the limit of large
skewness provided by the TFM. This new parametrization permits a much better
description of the FOM for convection in the atmosphere of the
Earth (GH2002).  We show that for their parametrization to be applicable to
the case of solar convection, a more realistic estimate for the skewnesses of
velocity and temperature fluctuations is required than that provided by the
TFM itself (Sect.~\ref{Sect_TFM}). 

 The parametrization of GH2002 requires the knowledge of the skewnesses and second-order 
 moments to compute FOM. These have to be provided
either by measurements, by another model, or by numerical simulations. In the 
present paper we do not aim to construct a complete model to compute 
these quantities, which is the goal of the Reynolds stress approach
(e.g., \citealt{Canuto92}; \citealt{Canuto98}). Rather, we aim to analyze 
the shortcomings of the TFM and suggest improvements using numerical
simulations of solar convection as a guideline. The conclusions drawn from
this analysis are used to derive a model for fourth-order moments in terms
of second-order moments that can be used in computations of solar $p$-mode
excitation rates.

 To proceed with the latter, we developed a formulation of
the TFM that takes the effects of turbulence in each flow into account. This
generalized TFM model (hereafter GTFM) is useful for both the superadiabatic
and adiabatic outer solar layers. This formulation can actually be applied in
other contexts than just the excitation of solar $p$~modes as long as the
convective system is composed of two flows.

The GTFM is more general and realistic than the TFM, but it requires the
knowledge of additional properties of both the turbulent upwards and downwards
flows. We choose to determine these properties by means of a plume model.
Turbulent plumes are created at the upper boundary of the convection zone,
where radiative cooling becomes dominant and where the flow reaches the stable
atmosphere. In this region the updrafts become cooler and stop their ascent. This
cooler flow is more dense than its environment and it triggers the
formation of turbulent plumes \citep{Stein98}. As shown by \cite{RZ95}, these
structures drive the dynamics of the flow; hence, to construct a closure
model, we study the plume dynamics developed by \cite{RZ95} (hereafter RZ95).
This makes it possible to build a {\em closure model with plumes (CMP)}, which
is valid in the solar quasi-adiabatic convective region. In a companion paper
\citep[][hereafter Paper~II]{Belkacem06b}, we generalize this one-point correlation 
model to a two-points correlation model and calculate the power injected into solar 
$p$~modes.

The paper is organised as follows: Sect.~2 introduces the TFM. 
Its validity is then tested with a 3D numerical simulation of the
uppermost part of the solar convection region. 
In Sect.~3, we extend the TFM formulation (GTFM) in order to take into account
turbulent properties of both upward and downward flows. We next investigate
the asymptotic limits of the GTFM. 
In Sect.~4, we construct the CMP with the help of the RZ95 plume model. We test
the validity of this model with results from the 3D simulation and show that
the use of the plume model limits the validity of the CMP to the quasi-adiabatic zone. 
The CMP is then used to obtain analytical
expressions for the third and fourth moments. Section~5 is dedicated to
discussions and conclusions.


%
\section{The two-scale mass-flux model}  \label{Sect_TFM}
\subsection{The model}

The TFM considers a convective medium composed of
 upward and downward flows that are horizontally averaged.
\begin{figure}[t]
\begin{center}
\includegraphics[height=6cm,width=9cm]{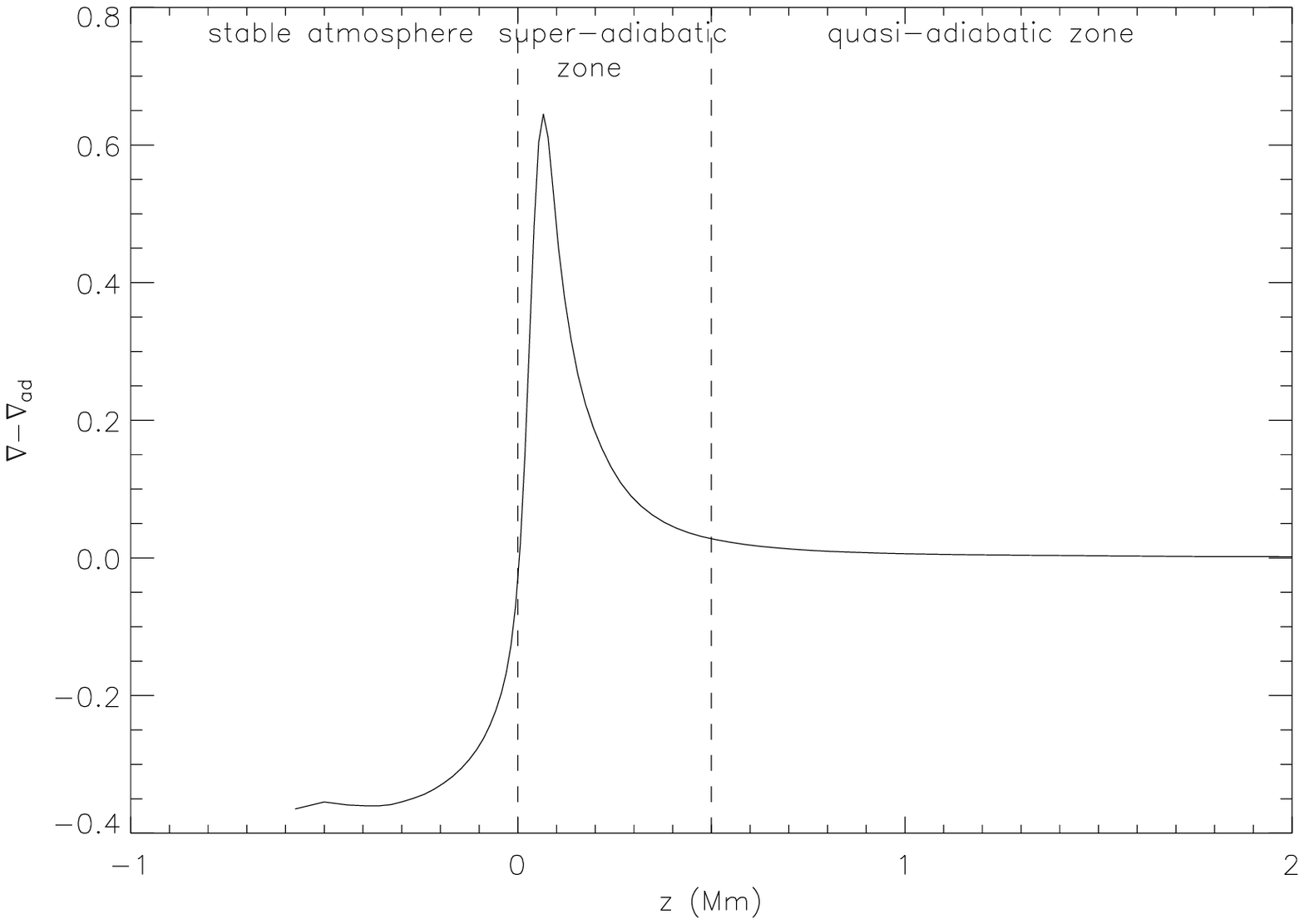}
\includegraphics[height=6cm,width=9cm]{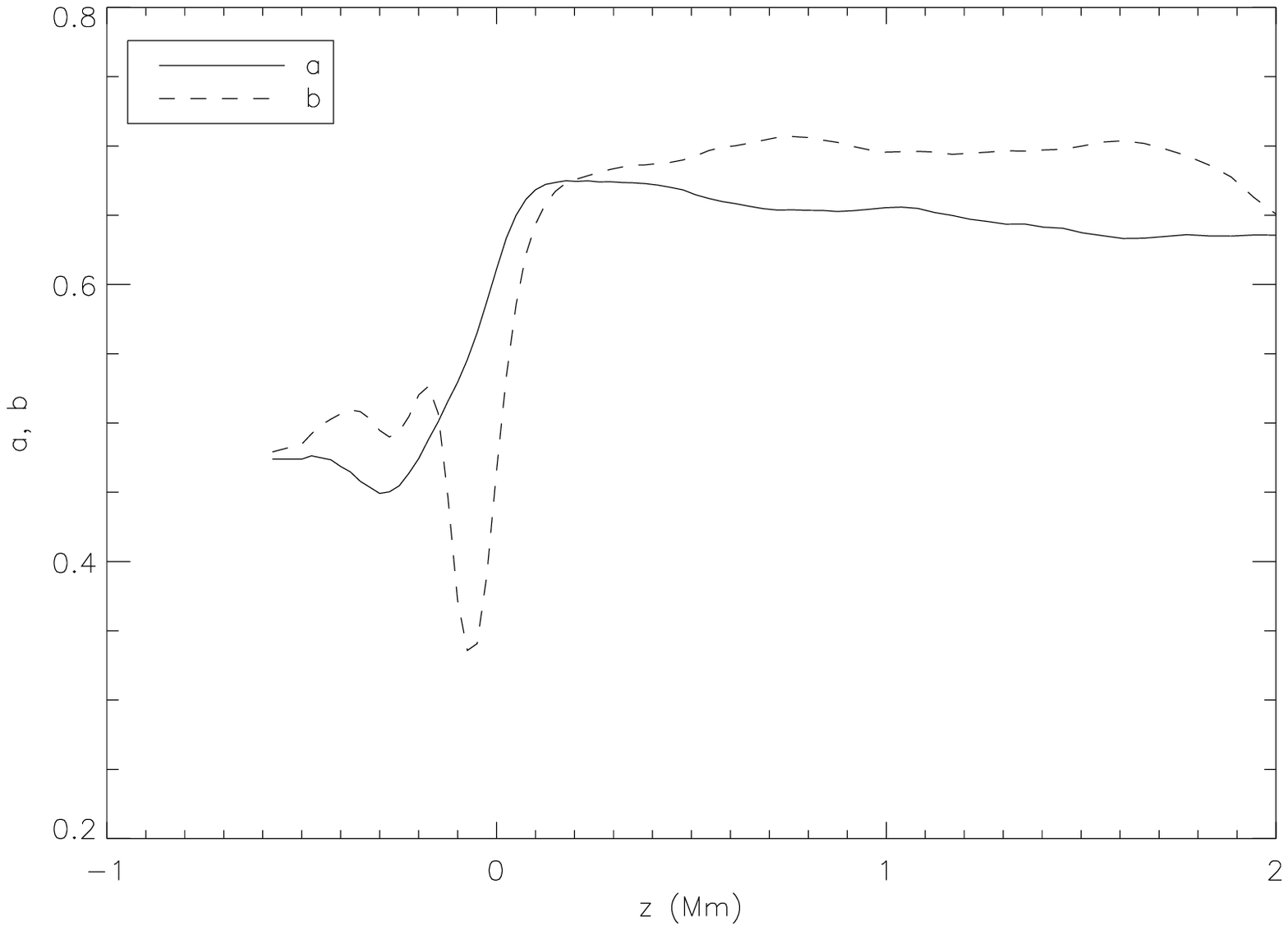}\\
\caption{On the top, the superadiabatic gradient ($\nabla-\nabla_{\rm ad}$) is
  plotted versus the depth ($z$). The reference depth ($z=0$~Mm) corresponds to the
  photosphere. At the bottom, the mean fractional area of the upflow ($a$) and
  the warm drafts ($b$) are given. To calculate these quantities the upflow and
  downflow are separated using the sign of $w'$ as a criterion. The same is done
  for the warm and cold drafts.}
\label{ab}
\end{center}
\end{figure}
The presence of two flows introduces the possibility of a non-zero
skewness for the moments of turbulent quantities when averages are done
globally over the whole system. The TFM was developed in order to take into
account this non-zero skewness.

Any averaged turbulent quantity $\phi$ can be split into two parts, one
associated with the updrafts and the other with the downdrafts:
\begin{eqnarray}
\label{phi_decomp}
   <\phi> ~ = ~ a <\phi>_u + \, (1-a) <\phi>_d \; ,
\end{eqnarray}
where $<>$ denotes ensemble spatial (in the horizontal plane) and time
averages. $<\phi>_u$ and $<\phi>_d$ are the averages for the upflow and
downflow, respectively. $a$ and $1-a$ are the mean fractional area occupied
by the updrafts and downdrafts, respectively (\citealt{Randall92};
\citealt{GH2002}; \citealt{Canuto98}).

Fluctuating\ quantities defined as $\phi' = \phi - <\phi>$ can be rigourously
written as: $<\phi'> = a ~<\phi'>_u+~ (1-a) ~<\phi'>_d$, where the subscripts
$u$ and $d$ are meant for upflow and downflow, respectively. For vertical velocity
fluctuations\ $w'$, one then writes:
\begin{eqnarray}
\label{decomp_1}
  <w'> ~=~ a ~<w'>_u + (1-a) ~<w'>_d \; .
\end{eqnarray}
GH2002 propose to make the same decomposition for temperature fluctuations\
($\theta'$); thus, hot and cold regions are considered separately.  This step
was motivated by the observation that for the case of atmospheric boundary
layer convection the characteristic horizontal scales of velocity and
temperature flucutations are different from each other and by the fact that
the plain mass flux average Eq.~(\ref{phi_decomp}) violates certain symmetries
between velocity and temperature flucutations. Indeed, hot and cold 
regions do not necessarily coincide with updrafts and downdrafts, respectively.
Hence, a second quantity ($b$), the mean fractional area occupied by warm
drafts, is introduced, and in most cases, $a \neq b$ (thus the name TFM). Then,
\begin{eqnarray}
   <\theta'>=b<\theta'>_h+(1-b)<\theta'>_c \; .
\end{eqnarray}

Furthermore, the TFM defines the velocity fluctuations\ inside the upflow
($w'_u$) and downflow ($w'_d$), respectively, as:
\begin{eqnarray}
\label{wu_wd}
   w'_u  &=& w_u-<w> \quad \mbox{\rm and} \quad w'_d=w_d-<w>.
\end{eqnarray}
Similarly, for the temperature fluctuations inside hot ($\theta'_h$) and
cold ($\theta'_d$) regions, respectively, one has
\begin{eqnarray}
\label{th_tc}
  \theta'_h &=& \theta_h-<\theta> \quad \mbox{\rm and} \quad
  \theta'_c=\theta_c-<\theta>.
\end{eqnarray}
 The quantities $w_u$, $w_d$, $\theta_h$, and $\theta_c$ are the averages
of velocity and temperature, respectively, over all updrafts ($w_u$),
downdrafts ($w_d$), hot ($\theta_h$) drafts, and cold ($\theta_c$) drafts.
Clearly, averages of the four fluctuating\ quantities in Eqs.~(\ref{wu_wd})
and~(\ref{th_tc}) do not vanish because the average of a quantity over the whole
flow differs from the average over one single (up or down, hot or cold) draft.

It is expected that the differences between the updrafts and downdrafts lead to
a probability distribution function (PDF) that is no longer symmetric with
respect to vanishing velocities and temperature differences. As the third-order 
moments ($<w'^3>$ and $<\theta'^3>$) vanish when the PDF is symmetric,
 their values provide a measure for the deviation from a symmetric
PDF. The skewnesses are defined as:
\begin{eqnarray}
\label{S_1}
  S_w = \frac{<w'^3>}{<w'^2>^{3/2}} \quad \mbox{\rm and} \quad
  S_{\theta} = \frac{<\theta'^3>}{<\theta'^2>^{3/2}} \; ,
\end{eqnarray}
respectively, for the vertical velocity and temperature fluctuations.
\begin{figure}[t]
\begin{center}
\includegraphics[height=6cm,width=9cm]{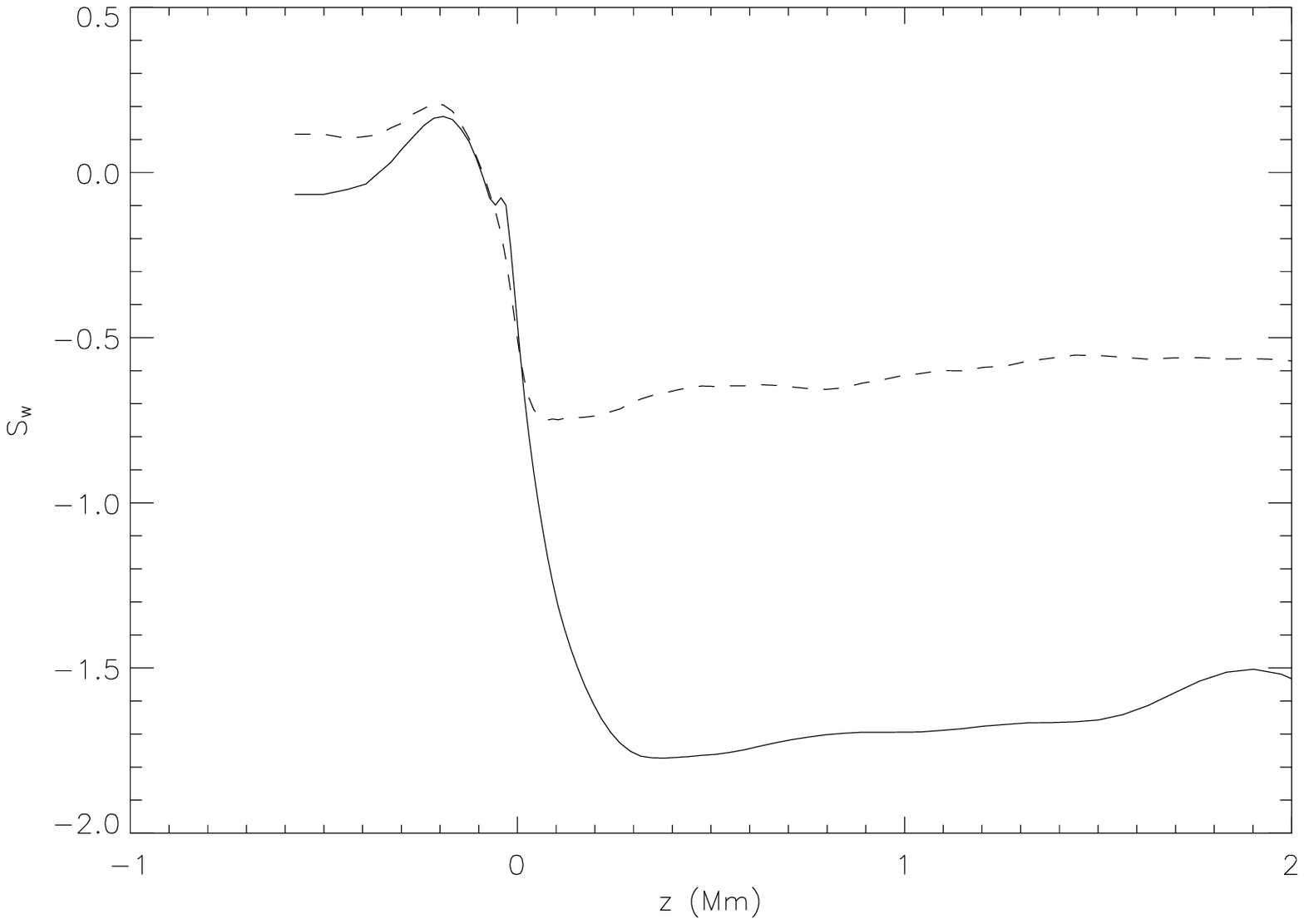}
\includegraphics[height=6cm,width=9cm]{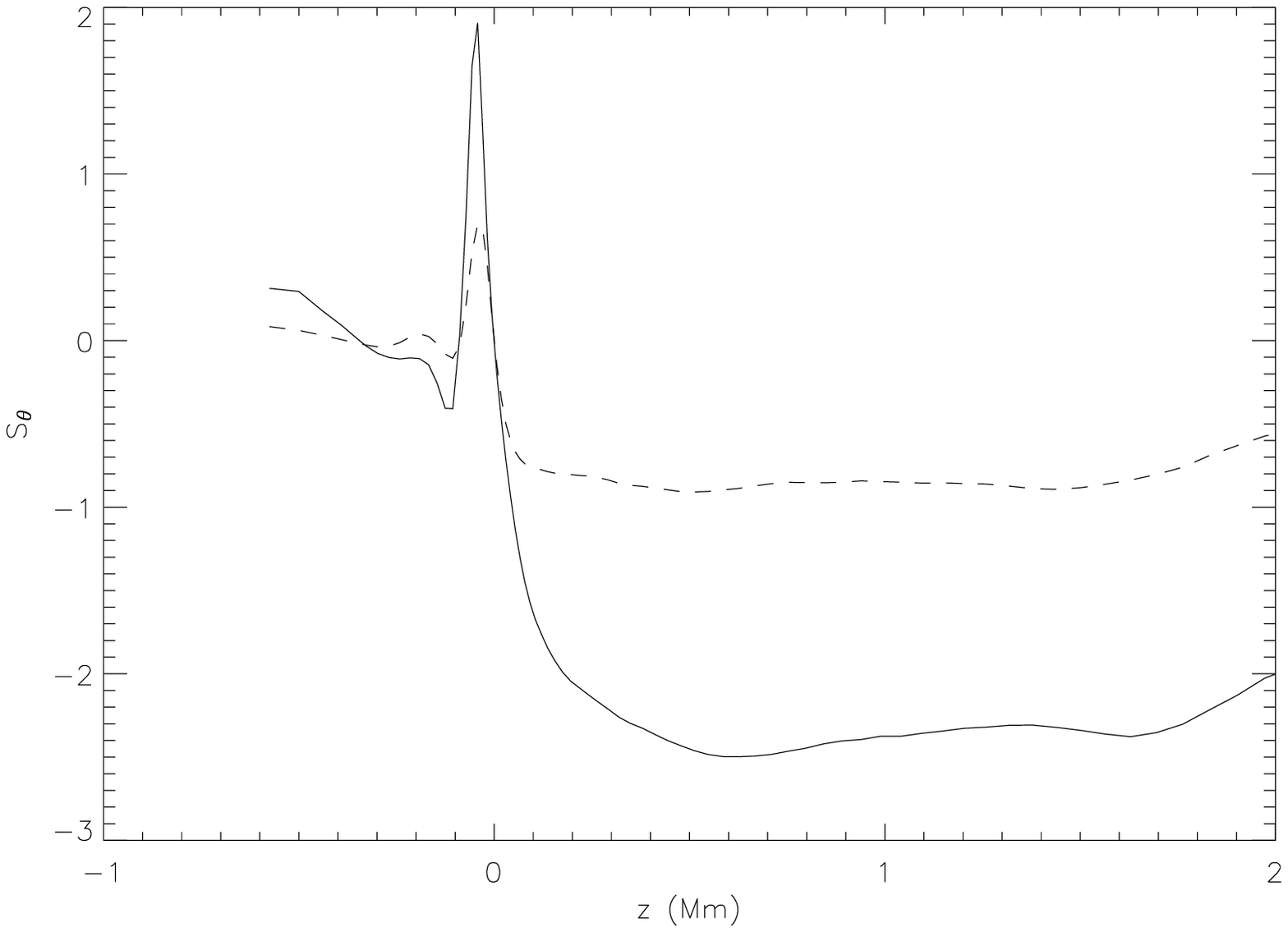}\\
\caption{The skewnesses $S_w$ (on the top) and $S_{\theta}$ (on the bottom) are
   plotted versus the depth ($z$). Solid lines represent direct
   calculation from the 3D numerical simulation (Eq.~(\ref{S_1})) and dashed
   lines represent the skewnesses calculated using the TFM model (Eq.~(\ref{S_2})).}
\label{Sw}
\end{center}
\end{figure}
In order to compute expressions for higher order moments in terms of
velocity and temperature fluctuations, Eqs.~(\ref{wu_wd}) and~(\ref{th_tc}), GH2002
followed \citet{Randall92}, using an additional simplifying approximation,
i.e.,\
\begin{equation}
\label{main_approx}
   <\phi^n> ~ \approx  ~ <\phi>^n \; ,
\end{equation}
where $\phi=\{ w'_{u,d},\theta'_{h,c}  \}$. This approximation neglects
the contributions of flucutations within the up- and downdrafts and differences in
temperature and velocity between the individual drafts.

Given this approximation and the known second-order moments, the TFM provides
third-order moments as follows (see GH2002):
\begin{eqnarray}
\label{no_inter_TOM}
  <w'^2 \theta'>  &=&  S_{w} <w'^2>^{1/2} <w' \theta'>  \\
  <w' \theta'^2>  &=&  S_{\theta} <\theta'^2>^{1/2} <w' \theta'>  \nonumber
\end{eqnarray}
and FOMs as:
\begin{eqnarray}
\label{no_inter}
  <w'^4>          &=&  (1+S^2_{w}) ~<w'^2>^2  \nonumber \\
  <\theta'^4>     &=&  (1+S^2_{\theta}) ~<\theta'^2>^2   \\
  <w'^3 \theta'>  &=&  (1+S^2_{w}) ~<w'^2> ~<w' \theta'>  \nonumber \\
  <w' \theta'^3>  &=&  (1+S^2_{\theta}) ~<\theta'^2> ~<w' \theta'> \; \nonumber .
\end{eqnarray}
The skewnesses $S_w$ and $S_{\theta}$ (Eq.~(\ref{S_1})) are related to $a$ and
$b$ through
\begin{equation}
\label{S_2}
  S_{w} = \frac{1-2a}{\sqrt{a(1-a)}} \quad \mbox{\rm and} \quad
  S_{\theta} =  \frac{1-2b}{\sqrt{b(1-b)}}
\end{equation}
 (GH2002, see also \citealt{Randall92} for the case of $S_{w}$).
 
In the following we consider only vertical-velocity moments.
Assuming $S_w=S_{\theta}=0$ in Eq.~(\ref{no_inter}) gives:
\begin{eqnarray}
  <w'^4> ~ = ~ <w'^2>^2 \; .
\end{eqnarray}
Such a result is not consistent with a quasi-normal (Gaussian) PDF. Indeed,
when $w'$ follows a normal distribution, then \citep{Lesieur97}:
\begin{eqnarray}
\label{Def_QNA}
  S_w = S_{\theta} = 0 \quad \mbox{and} \quad <w'^4> ~ = ~ 3 <w'^2>^2 \; .
\end{eqnarray}
  GH2002 found that the two-scale mass-flux average,
Eqs.~(\ref{no_inter_TOM}),-,(\ref{S_2}), underestimates both skewness and fourth-order 
moments as measured by aircraft data for planetary boundary layer
convection (see their Figs.~4 and~7). To account for the omitted contributions
from fluctuations within and between the up- and downdrafts, they suggested generalizing the
TFM by building the fourth-order moments as an interpolation between two
asymptotic regimes:
\begin{itemize}
\item $S_w=0$, assuming the quasi-normal approximation (QNA) limit that is
      valid for a Gaussian PDF, and
\item $S_w >> 1$, the large skewness limit (GH2002).
\end{itemize}
GH2002 hence proposed:
\begin{eqnarray}
\label{inter}
  <w'^4>          &=&  3~(1+\frac{1}{3}S^2_{w}) ~<w'^2>^2  \nonumber \\
  <\theta'^4>     &=&  3~(1+\frac{1}{3}S^2_{\theta}) ~<\theta'^2>^2   \\
  <w'^3 \theta'>  &=&  3~(1+\frac{1}{3}S^2_{w}) ~<w'^2> ~<w' \theta'>  \nonumber \\
  <w' \theta'^3>  &=&  3~(1+\frac{1}{3}S^2_{\theta}) ~<\theta'^2> ~<w' \theta'> \nonumber \; .
\end{eqnarray}
Corresponding expressions for other FOMs ($<w'^2 \theta'^2>$ and those including
horizontal velocities) can be found in \citet[][hereafter GH2005]{GH2005}. 

\subsection{Validation with a 3D numerical simulation of the solar
            external layers}
\label{model_numeric}

\begin{figure}[t]
\begin{center}
\includegraphics[height=6cm,width=9cm]{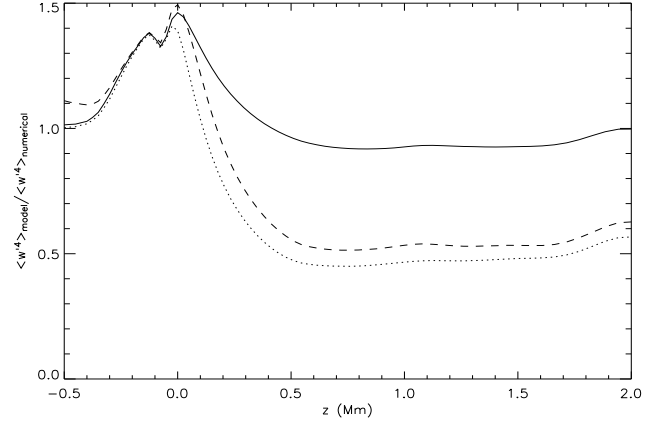}\\
\caption{Fourth-order moment ($<w'^4>$) as a function of depth ($z$) normalized
   to the FOM, as calculated directly from the simulation. The solid
   line denotes the moment calculated using Eq.~(\ref{inter}) with $S_w$ taken
   directly from the simulation; the dashed line shows the result if $S_w$
   is instead taken from Eq.~(\ref{S_2}), as in the TFM case; and the dotted
   line is the QNA (Eq.~(\ref{w4_QNA})). Equations.~(\ref{inter}) and~(\ref{w4_QNA})
   involve second-order moments that are computed using the numerical
   simulation.}
\label{w4_1}
\end{center}
\end{figure}
We consider the uppermost part of the solar turbulent convection. Turbulent
plumes are known to exist within this region \citep{Cattaneo91,Stein98}. Here,
we test the validity of the TFM  using 3D numerical simulations of these upper
solar layers. The geometry is plane-parallel with a physical size of
$6 \, \mbox{\rm Mm}\times 6 \, \mbox{\rm Mm} \times 3 \, \mbox{\rm Mm}$.
The upper boundary corresponds to a convectively stable atmosphere and the
lower one to a quasi-adiabatic convection zone. 
The 3D simulations used in this work were obtained with Stein \& Nordlund's 3D
numerical code \citep{Stein98}.
Two simulations with different spatial grids were considered:
$253 \times 253 \times 163$ and $125 \times 125 \times 82$.

 Averages  and moments of the velocity and temperature fluctuations were
computed in a two-stage process: \\
$a$ is given as the number of grid points
per layer with upwards directed vertical velocity divided by the total
number of points in that layer. The instantaneous value of
$b$ is obtained in a similar manner, comparing the temperature at a given point in a layer with its
horizontal average. Moments related to updrafts were obtained from
horizontal averaging, using only those grid points at which vertical
velocity was directed upwards at the given instant in time, and likewise,
quantities related to downdrafts were obtained from horizontal averaging
using only those grid points at which vertical velocity was directed
downwards. 
In a second step, time averages were performed over a sufficiently
long period of time such that averages no longer depended on the
integration time beyond a few percent. \\

{\it ---  Calculation of the skewnesses:}
Computations of the mean fractional area of the upflow $(a)$ and downflow
$(1-a)$, as well those of the warm $(b)$ and cold $(1-b)$ drafts from the
numerical 3D simulations (Fig.~\ref{ab}), show that the upper part of the solar
convection zone can be divided into three parts: the stable atmosphere, the
superadiabatic zone, and the quasi-adiabatic zone. 
In the convectively stable atmosphere ($z < 0$~Mm, where $z=0$ is approximately
 at the bottom of the photosphere and $z=-0.5$~Mm is the uppermost boundary of
the simulation), there are no asymmetric motions. 
In the superadiabatic zone ($0 < z < 0.5$~Mm), from the top downwards, the
departure from symmetry for the flows strongly increases (Fig.~\ref{ab}), and
the skewnesses, $S_w$ and $S_{\theta}$, significantly differ from zero
(Fig.~\ref{Sw}). Hence, one must expect a non-negligible departure from the QNA,
which is explained by radiative cooling creating turbulent plumes. In the
quasi-adiabatic zone, plumes have already been formed and no additional
asymmetry is therefore created. Hence, the asymmetry remains large and constant
($a \approx b \approx 0.7$) and the skewnesses show a constant departure
from $S_w=S_{\theta}=0$.

The last two regions are of interest in this work because both 
show a departure from the quasi-normal PDF in terms of fluctuating
vertical velocity and temperature. 
The comparison of the above numerical results with the results from the
classical TFM (Eq.~(\ref{no_inter})) and the TFM model (Fig.~\ref{Sw}) shows that
Eq.~(\ref{S_2}) fails to reproduce the behaviour of the skewnesses from the
3D simulation (as was also found by \citealt{GH2002} for convection in the
atmosphere of the Earth, see their Fig.~4).

{\it ---  Detailed comparison of a fourth-order moment:}
The GH2002 interpolation relation Eq.~(\ref{inter})
    combined with the TFM relation for skewness, Eq.~(\ref{S_2}),
shows only a slight improvement of the QNA description for the
FOM $<w'^4>$, when compared to the numerical result (Fig.~\ref{w4_1}).

To conclude, it seems that a physical process is missing in the quasi-adiabatic
convective zone. To explain such a disagreement between the numerical results
and the TFM, we must come back to its main approximation (see
Eq.~(\ref{main_approx})). For $n=2$, Eq.~(\ref{main_approx}) yields:
\begin{equation}
  <w'^2> - <w'>^2 ~\approx ~0 \quad <\theta'^2> - <\theta'>^2 \approx 0 \; .
\end{equation}
Hence, the TFM assumes that the variances of the fluctuations of vertical
velocity and temperature within and among individual drafts vanish, and the detailed turbulent nature of the
flows themselves does not have to be taken into account. In order to compensate for the
shortcoming of Eq.~(\ref{no_inter}) and thus the consequences of the approximation Eq.~(\ref{main_approx})
on the model predictions, \cite{GH2002} proposed a more general interpolation relation
(Eq.~(\ref{inter})) that uses Eq.~(\ref{no_inter}) only for one of
two asymptotic limits.

As seen above, Eq.~(\ref{S_2}) fails to describe the numerical results. The
question therefore is whether the interpolated relation (Eq.~(\ref{inter})) is
still valid, provided a correct value for the skewness is used. Hence, we assess
the validity of Eq.~(\ref{inter}) by inserting the value of $S_w$ directly given
by the 3D numerical result. The result is shown in Fig.~\ref{w4_1} as well. This
is the model that \cite{GH2002} proposed to be used instead of the
TFM itself and its associated relation for the skewnesses, Eq.~(\ref{S_2}).
We obtain an accurate description of the FOM $<w'^4>$ in the
quasi-adiabatic region, but not in the superadiabatic zone, where the
interpolated relation does not seem well adapted (cf. KR2006 for a more detailed
discussion).

%
\section{The generalized two-scale mass-flux model}
\subsection{Theoretical formulation}   \label{Sect_GTFM_formulation}

Here we remove the approximation of Eq.~(\ref{main_approx}) and instead consider the
exact expression:
\begin{equation}
\label{exact_decomp}
   <w'^n> = a <w'^n>_u + ~(1-a) <w'^n>_d \; .
\end{equation}
Our main idea is to separate the effect of the skewness introduced by the
presence of two flows from the effect of the turbulence that occurs in each
individual flow. We note that in a geophysical context \cite{Siebesma95}
and \cite{Petersen99} studied the transport properties of classical
mass-flux models that also involved a separation of large-scale and turbulent
components. Here, we start from the more recent viewpoint of the TFM by
\cite{GH2002} and \citet{GH2005}, which takes into account that updrafts and
downdrafts are not strictly correlated with hot and cold drafts, respectively.
As a first step we define the intrinsic fluctuations within one of the flows as:
\begin{eqnarray}
\label{int_fluct}
  \tilde{w}'_{j} = w_{j} - <w>_j \; ,
\end{eqnarray}
where $j=\{u, d\}$. They are fluctuations with vanishing averages.
To express $w'_{j}$ in terms of $\tilde{w}'_{j}$ (Eq.~(\ref{wu_wd})), we write:
\begin{equation}
     w'_{j} ~= ~\tilde{w}'_{j}~+~<w>_{j}~-~<w> \; .
\end{equation}
Applying the decomposition of Eq.~(\ref{phi_decomp}) to $<w>$ in the above
expression yields:
\begin{eqnarray}
\label{proper}
   w'_{u} &=& \tilde{w}'_{u}~+~(1-a) ~ \delta w \nonumber \\
   w'_{d} &=& \tilde{w}'_{d}~-~a ~ \delta w
\end{eqnarray}
with
\begin{equation}
\label{del_w}
   \delta w= <w>_u-<w>_d= |<w>_u|+|<w>_d|,
\end{equation}
because $<w>_u \, > 0$ and $<w>_d \, < 0$.\\
 Inserting Eq.~(\ref{proper}) into Eq.~(\ref{exact_decomp}) for $n=2, 3$ yields:
\begin{eqnarray}
\label{SOM}
   <w'^2> &=& a(1-a) \, \delta w^2 \nonumber \\
          &+& a<\tilde{w}'^2>_u+~(1-a)<\tilde{w}'^2>_d
\end{eqnarray}
\begin{eqnarray}
\label{TOM}
   <w'^3> &=& a(1-a)(1-2a) \, \delta w^3 \nonumber \\
	       &+& a<\tilde{w}'^3>_u+~(1-a)<\tilde{w}'^3>_d \nonumber \\
               &+& 3a(1-a)\Big[<\tilde{w}'^2>_u-<\tilde{w}'^2>_d\Big]\delta w \; .
\end{eqnarray}
The third-order moment (Eq.~(\ref{TOM})), which is related to the skewness (see
Eq.~(\ref{S_1})), is composed of four contributions:
\begin{itemize}
\item the first term is the expression derived by \cite{GH2002}. It is a measure
      of the skewness introduced by the presence of two flows.
\item the second and third terms represent the asymmetry of the PDF within each
      flow induced by turbulence.
\item the fourth term measures the difference of the fluctuating velocity
      dispersion. Hence, if one of them is larger than the other, the PDF becomes
      asymmetric.
\end{itemize} 
The description of the turbulence in individual flows that has been
neglected in the TFM is included in the present formulation through the
last three terms in Eq.~(\ref{TOM}).

We next focus on the fourth-order moment $<w'^4>$, which is of interest in the
context of stochastic excitation of solar $p$~modes (see Paper II). Then setting
 $n=4$ in Eq.~(\ref{exact_decomp}), we have:
\begin{eqnarray}
\label{FOM}
<w'^4> &=& a(1-a)(1-3a+3a^2) \; \delta w^4 \nonumber \\
            &+& 6a(1-a)\Big((1-a)<\tilde{w}'^2>_u+a<\tilde{w}'^2>_d\Big)\delta w^2 \nonumber \\
	    &+& 4a(1-a)\Big(<\tilde{w}'^3>_u-<\tilde{w}'^3>_d\Big)\delta w \nonumber \\
	    &+& a<\tilde{w}'^4>_u + ~(1-a)<\tilde{w}'^4>_d \; .
\end{eqnarray}
We stress that the TFM is recovered from the present generalized formulation
when proper fluctuations (i.e., turbulence) within and among the individual drafts are
neglected, i.e., $<\tilde{w}'^n>=0$. 

The same decomposition can be performed in terms of temperature fluctuations.
As the calculation is symmetrical in $w',a$ and $\theta',b$, we hence have:
\begin{eqnarray}
\label{theta_STFOM}
<\theta'^2> &=& b(1-b) \, \delta\theta^2 \nonumber \\
            &+& b<\tilde{\theta}'^2>_h+(1-b)<\tilde{\theta}'^2>_c \nonumber \\
<\theta'^3> &=& b(1-b)(1-2b) \, \delta\theta^3 \nonumber \\
            &+& b<\tilde{\theta}'^3>_h+~(1-b)<\tilde{\theta}'^3>_c \nonumber \\
	    &+& 3b(1-b)\Big[<\tilde{\theta}'^2>_h-<\tilde{\theta}'^2>_c\Big] ~ \delta\theta \nonumber \\
<\theta'^4> &=& b(1-b)(1-3b+3b^2) \; \delta\theta^4 \nonumber \\
            &+& 6b(1-b)\Big((1-b)<\tilde{\theta}'^3>_h+ b<\tilde{\theta}'^2>_c\Big) ~ \delta\theta^2 \nonumber \\
            &+& 4b(1-b)\Big(<\tilde{\theta}'^3>_h-<\tilde{\theta}'^3>_c\Big) ~ \delta\theta \nonumber \\
	    &+& b<\tilde{\theta}'^4>_h + (1-b)<\tilde{\theta}'^4>_c \; .
\end{eqnarray}
The next step consists of the derivation of the cross terms
$<w'\theta'>,<w'^2\theta'^2>, <w'^2\theta'> and <w'\theta'^2>$;
it is convenient to  define the coefficients $a_{uh}, a_{uc}$ so as to take into
account the four types of flow (see also GH2005):
\begin{itemize}
\item warm updraft, $a_{uh}$
\item cold updraft, $a_{uc}=a-a_{uh}$
\item warm downdraft, $a_{dh}=b-a_{uh}$
\item cold downdraft, $a_{dc}=1-b-a_{uc}$
\end{itemize}
Expressions for the third and fourth cross-correlation moments are given
in Appendix~\ref{Sect_app_A}. 

The generalized TFM has the advantage of isolating the skewness introduced
by the two flows (as measured by $S_w$ and $S_{\theta}$ in Eq.~(\ref{S_2})) from
the effects of turbulence in each of the flows (as measured for instance by the
two terms $\tilde{w}'^2_d$ and $\tilde{w}'^2_u$). 
 The GTFM allows us to take the effects of turbulence into account.
 We note
that a small value of the kurtosis can occur only if proper fluctuations
lead to negligibly small deviations from the root mean square 
average. Such a flow pattern consisting of clearly defined up- and 
downflows as well as hot and cold areas with a kurtosis $K_w \gtrsim 1$ 
can be considered as representing a {\em quasi-laminar state}. 
 We
stress that for the quasi-laminar case, Eq.~(\ref{no_inter}) remains exact; thus the
kurtosis becomes:
\begin{eqnarray}
\label{Kurt_laminar}
K_w = \frac{<w'^4>}{<w'^2>^2} = (1+S_w^2) \; \mbox{  with} \; S_w=\frac{1-2a}{\sqrt{a(1-a)}} \; .
\end{eqnarray}
For $a=0.5$, one obtains $K_w=1$, which is far from the value for a Gaussian PDF
($K_w=3$). 
To take into account turbulence within the up- and downdrafts, one can
use Eq.~(\ref{inter}) (see Sect.~\ref{model_numeric}) with the skewness
$S_w=<w'^3>/<w'^2>^{3/2}$ from the GTFM. In this case we obtain:
\begin{equation}
\label{kurtosis}
K_w = 3(1+\frac{1}{3}S_w^2) \; .
\end{equation}
This implies that a (moderately small) non-vanishing skewness will make the
value of $K_w$ closer to three than in the quasi-laminar case. In the solar
case, in the quasi-isentropic zone $S_w^2 \approx 4$ (Fig.~\ref{Sw}), hence
$K_w \approx 3+4/3$.
In the physical picture underlying Eq.~(\ref{inter}), turbulence prevents 
 the PDF from being too far from a Gaussian one
($K_w \to 3$).

We notice that one important source of turbulence that can
be considered responsible for at least part of the fluctuations in a
draft --- in addition to those created by the radiative processes on top
of the convection zone --- is related to shearing stresses between the
up- and downdrafts. However, the investigation of the sources of turbulence 
is beyond the scope of the present work.
Those mechanisms certainly play an important role in both the
small scale velocity and the thermal fluctuations. Their study is
definitely desirable in the future. 
One should also note that the splitting approach of the GTFM is valid and can be used for
any convective system, provided that it is composed of two flows. 
 As it is unclosed, it must be seen as a good basis for building a closure model.

\subsection{Asymptotic limits}
\label{limits}

In the following, we study the asymptotic limits of the GTFM, focusing on the
fourth-order moment $<w'^4>$. The standard mass flux model is easily
recovered when setting the proper moments to zero: $<\tilde{w'}^n>=0$ in
Eqs.~(\ref{SOM}),--,(\ref{FOM}). The same holds for the TFM,
Eqs.~(\ref{no_inter_TOM}),--,(\ref{no_inter}), which is recovered, if in addition
$<\tilde{\theta'}^n>=0$ in Eq.~(\ref{theta_STFOM}) (cf.\ Eqs.~(7) and (8) in
\citealt{GH2002}). We now turn to the QNA limit and the limit for large skewness,
which are more interesting as they are used by \cite{GH2002} and \citet{GH2005}
in order to corroborate the interpolation formula Eq.~(\ref{inter}).

\subsubsection{The quasi-normal limit}

To obtain the QNA (Eq.~(\ref{Def_QNA})), it is necessary that $S_w=0$, but it is
not sufficient. In fact, a vanishing skewness only shows that the PDF is
symmetric, \emph{but not that the PDF is Gaussian}. 
Further conditions are necessary:
\begin{itemize}
\item the moments must have zero mean, which implies
      $|<w>_u|=|<w>_d|=0$ from Eq.~(\ref{decomp_1}) and Eq.~(\ref{proper});
\item for the QNA to apply to the whole system, one must
      assume that the QNA is valid for each flow;
\item we must also assume that $a=0.5$;
\item the turbulent pressure must be the same in the upflow and downflow.
      Otherwise the skewness ($S_w$) is different from zero, according to
      Eq.~(\ref{TOM}), and the consequence is an asymmetric PDF, which is not
      consistent with the quasi-normal assumption. So the condition
      $<\tilde{w}'^2>_u=<\tilde{w}'^2>_d$ is required.
\end{itemize}
Then starting with Eq.~(\ref{FOM}), we find:
\begin{eqnarray}
<w'^4> &=& a \, <\tilde{w}'^4>_u+~(1-a) \, <\tilde{w}'^4>_d \nonumber \\
       &=& \frac{3}{2}<\tilde{w}'^2>_u^2+\frac{3}{2}<\tilde{w}'^2>_d^2 \nonumber \; .
\end{eqnarray}
and finally
\begin{equation}
\label{w4_QNA}
   <w'^4>~=~3<w'^2>^2 \; .
\end{equation}
which is the expression for the fourth order moment in the QNA. 
Note that the TFM (Sect.~\ref{Sect_TFM}) is unable to properly recover the QNA.
Within the GTFM the QNA results from two terms, $<\tilde{w}'^4>_u$ and
$<\tilde{w}'^4>_d$, which are related to the intrinsic turbulence in each
flow, but these are neglected in the TFM. This example also demonstrates
that for a convective flow, the deviation of a PDF from a Gaussian one cannot be
modelled by the TFM without further modifications of that model (even if $a=0.5$).

\subsubsection{The large skewness limit}

\cite{GH2005} have shown that the TFM must be recovered when considering
a convective system with large skewness. Then, for $S_w >> 1$, the
expression for $<w'^4>$ in Eq.~(\ref{inter}) becomes:
\begin{eqnarray}
  <w'^4> \approx S_w^2 <w'^2>^2 \; .
\end{eqnarray}
The large skewness limit physically corresponds to either $a \approx 1$ or
$a \approx 0$. Indeed, it means that one of the two f\/lows\ dominates over
the other one in terms of mean fractional area in the horizontal plane. Thus,
due to conservation of mass, the mean vertical velocity becomes large such that
$\delta w >> 1$~m~s$^{-1}$ in the solar case (see Sect.~\ref{CMP},
Eq.~(\ref{conserv_mass})).

In Eq.~(\ref{FOM}), the term proportional to $\delta w^4$, which measures the
 effects introduced by an asymmetric flow, dominates and leads to the
TFM expression for the fourth-order moment $<w'^4>$:
\begin{equation}
   <w'^4> = a(1-a) \Big( a^3 + (1-a)^3\Big) \delta w^4 \; .
\end{equation}
\cite{GH2002} demonstrated that this expression leads to the relation:
\begin{eqnarray}
<w'^4> = (1+S^2) <w'^2>^2 \approx S^2 <w'^2>^2 \; \mbox{for} \; S >> 1 \; ,
\end{eqnarray}
where, as in Eq.~(\ref{S_2}), $S=(1-2a) / \sqrt{a(1-a)}$.  The same would
result if the exact function $S_w$ were taken in this limit instead of its
approximation, Eq.~(\ref{S_2}).

Hence, the GTFM enables us to show that the asymptotic limits used by
\cite{GH2002} to motivate the interpolated expressions for the FOMs
(Eq.~(\ref{inter})) are limiting cases for a flow that consists of a coherent
part with two components (up- and downdrafts), which themselves are subject
to turbulence (cf.\ the discussion of the GH2002 model in KR2006).
In Sect.~\ref{model_numeric} we have shown, using the 3D numerical simulation,
that this interpolation is valid provided the skewness is taken directly from
the 3D simulation. This property can be understood using the GTFM, as it
permits us to obtain the different ingredients of the interpolation formula
of \cite{GH2002} from Eq.~(\ref{FOM}) and the individual contributions to
Eq.~(\ref{FOM}), can be analyzed using numerical simulations.

%
\section{The closure model with plumes}
\label{CMP}

Section~\ref{model_numeric} confirmed the conclusion by KR2006 that the
interpolated relations in Eq.~(\ref{inter}) proposed by \cite{GH2002} could be
adapted for the solar case provided that the skewnesses are appropriately
calculated. Using the GTFM to model skewnesses, Eq.~(\ref{TOM}) shows that the
skewness $S_w$, for instance, depends on six quantities:
$\delta w,<\tilde{w}'^3>_{u,d}$, $<\tilde{w}'^2>_{u,d}$, and $a$. As shown
below, some of the terms in $S_w$ turn out to be negligible in the
quasi-adiabatic convective region because plumes are more turbulent in the 
downflow than in the upflow \citep{Stein98}. The remaining dominant terms are
modelled hereafter by a plume model \citep{RZ95} in the quasi-adiabatic
convective region, where the CMP is valid.

\begin{figure}[t]
\begin{center}
\includegraphics[height=6cm,width=9cm]{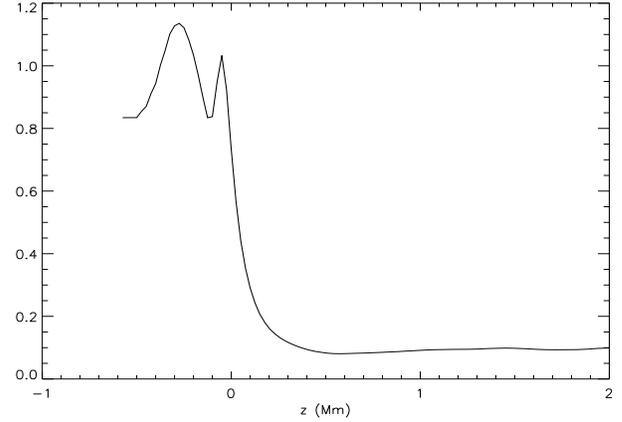}\\
\caption{Second-order moment of the upflow over that of the downflow 
  ($<\tilde{w}'^2>_u/<\tilde{w}'^2>_d$) as a function of depth, calculated
  directly from the simulation. Upflow and downflow are determined according
  to the sign of $w'$.}
\label{ordre2}
\end{center}
\end{figure}

\begin{figure}[t]
\begin{center}
\includegraphics[height=6cm,width=9cm]{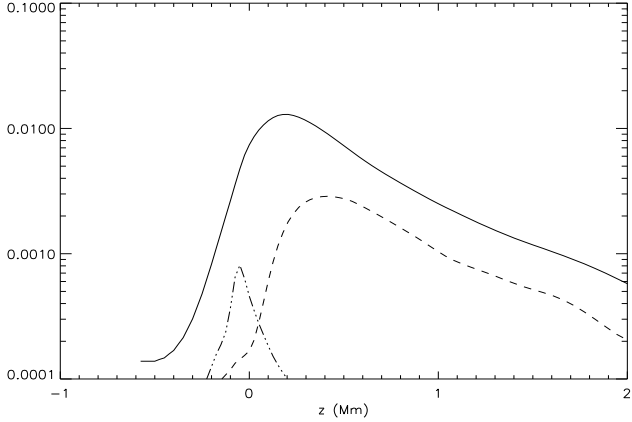}\\
\caption{The terms $3a(1-a)<\tilde{w}'^2>_d \delta w$ (solid line),
    $a<\tilde{w}'^3>_u$ (dot-dot-dot-dashed line), and $(1-a)<\tilde{w}'^3>_d$
    (dashed line) are plotted versus the depth ($z$). From Eq.~(\ref{TOM}),
    the dominant terms remains $3 a (1-a) <\tilde{w}'^2>_d \delta w$. This
    justifies the assumptions that the terms involving third-order moments
    can be neglected in the quasi-adiabatic zone.}
\label{test_3}
\end{center}
\end{figure}

\subsection{Turbulence in upflows and downflows}
\label{turb}

In Fig.~\ref{ordre2}, we compare the second-order moments of both flows.
These quantities are of the same order of magnitude in the upper part, above the
photosphere. From the photosphere, the ratio $<\tilde{w}'^2>_u/<\tilde{w}'^2>_d$
then sharply decreases, with increasing depth ($z$). Hence, contributions to the
skewness ($S_w$), involving $<\tilde{w}'^2>_u$ (Eqs.~(\ref{SOM}) and Eq.~(\ref{TOM}))
can be neglected in comparison with those involving $<\tilde{w}'^2>_d$ in the
quasi-adiabatic part of the convection zone. The third-order moments
$<\tilde{w}'^3>_{d}$ and $<\tilde{w}'^3>_{u}$ can also be discarded (see
Fig.~\ref{test_3}) because their contributions are negligible.

The skewness $S_w$ then becomes:
\begin{equation}
\label{S_w_CMP_compl}
  S_{w}~=~\frac{a(1-a)}{<w'^2>^{3/2}}
                  \Big((1-2a) \delta w^2-3<\tilde{w}'^2>_d\Big) \, \delta w \; ,
\end{equation}
where $\delta w$ is given by Eq.~(\ref{del_w}). Hence, only
$<\tilde{w}'^2>_{d}$ and $\delta w$ remain to be modelled. Similarly, the
3D calculations show that the cool medium is more turbulent than the hot
one and that third-order moments for the temperature fluctuations can be neglected.
Then the expression for $S_{\theta}$ becomes:
\begin{eqnarray}
\label{S_teta_CMP_compl}  
  S_{\theta}=b(1-b) \, \frac{1} {<\theta'^2>^{3/2}}
     \Big((1-2b) \delta\theta^2 -3<\tilde{\theta}'^2>_c\Big)~ \delta\theta \; ,
\end{eqnarray}
where the quantities $\delta\theta= <\theta>_h - <\theta>_c$ and
$<\tilde{\theta}'^2>_c$ must be modelled.

 Note that in the QNA limit $\delta w = 0$, so that for the expression Eq.~(\ref{S_w_CMP_compl}), 
$S_w=  0$, and according to Eq.~(\ref{inter}), \\$<w'^4> = 3 <w'^2>^2$.\\ 
However, because we have assumed $<\tilde{w}'^2>_u ~ << ~<\tilde{w}'^2>_d$
 when deriving the expression $S_w$, rigourously speaking, 
  $S_w$ does not tend correctly to zero in the QNA limit. Such an expression
  therefore cannot be used in the case of a near QNA regime. In our case, we
  have shown in Sect.~\ref{model_numeric}  that the medium 
  is far from the QNA limit in the quasi-adiabatic zone, and hence the
  expression Eq.~(\ref{S_w_CMP_compl})  can be safely used.

To proceed further, $<\tilde{w}'^2>_d$ and $<\tilde{\theta}'^2>_c$ are written
in a more suitable form. We neglect $<\tilde{w}'^2>_{u}$ in Eq.~(\ref{SOM}) for
$<w'^2>$, and $<\tilde{\theta}^2>_{h}$ in Eq.~(\ref{theta_STFOM}) for $<\theta'^2>$.
This yields:
\begin{eqnarray}
\label{ordre_2_CMP}
  <w'^2> &=& a(1-a)~\delta w^2~+~(1-a)<\tilde{w}'^2>_d \nonumber \\
  <\theta'^2> &=& b(1-b)~\delta \theta^2~+~(1-b)<\tilde{\theta}'^2>_c \; .
\end{eqnarray}
We then derive expressions for $<\tilde{w}'^2>_d$ and $<\tilde{\theta}'^2>_c$
in terms of $<w'^2>$, $\delta w$, and $<\theta'^2>$, $\delta \theta$,
respectively (see Eq.~(\ref{ordre_2_CMP})).
Inserting them into Eqs.~(\ref{S_w_CMP_compl}) and
(\ref{S_teta_CMP_compl}), one then obtains:
\begin{eqnarray}
\label{Sw_eff}
  S_{w}=  \frac{1}{<w'^2>^{3/2}}~ a~  \Big((1-a)(1- 5a)\delta w^2-  3 <w'^2> \Big)~ \delta w
\end{eqnarray}
and
\begin{eqnarray}
\label{S_teta_CMP}
  S_{\theta} = \frac{1} {<\theta'^2>^{3/2}} ~b~ \Big((1-b) (1-5b) \delta\theta^2 
                          - 3  <\theta'^2> \Big)~ \delta\theta \; .
\end{eqnarray}
We assume that the second-order moments ($<w'^2>$ and $<\theta'^2>$) are known.
In the present work, they are computed from the 3D numerical simulation. 
 In principle, they could also be taken from a convection model such as
the mixing-length theory. The
last step then is to determine $\delta w$ and $\delta \theta$ (as well
as $a$ and $b$). As $\delta w$ is the difference between the mean velocities
of upward and downward flows, it is possible to model it by means of a plume
model. This approach is also used to determine $\delta \theta$.

\subsection{The plume model}
\subsubsection{Determination of $\delta w$}

\begin{figure}[t]
\begin{center}
\includegraphics[height=6cm,width=9cm]{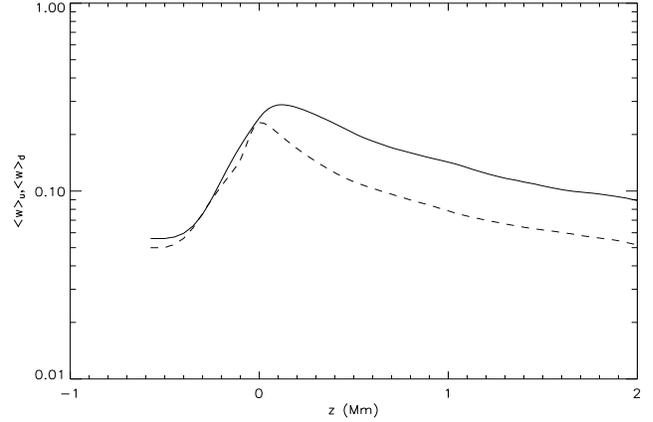}\\
\caption{Mean velocity profile of the upflow (dashed line) and downflow (solid
    line) as a function of the depth. Note that the peak at $z=0.1$~Mm
    corresponds to the maximum turbulent pressure. The use of power laws limits
    the validity of the CMP to the quasi-adiabatic zone, as is implied by the
    deviation of the profiles from power laws in the superadiabatic region.}
\label{auto_sim}
\end{center}
\end{figure}

We use the model of plumes developed by \cite{RZ95}. The plume is considered
in an axisymmetric geometry with a Gaussian horizontal profile for the vertical 
velocity ($w_d$), the
fluctuations of enthalpy ($\delta h$), and density ($\delta \rho$) such that
\begin{eqnarray}
        w_d(r,z) &=& V(z) \; \; \exp(-r^2/b_p^2)\; , \nonumber \\
\delta \rho(r,z) &=& \Delta  \rho(z) \, \exp(-r^2/b_p^2)\; , \rm{and} \nonumber \\
   \delta h(r,z) &=& \Delta h(z) \, \exp(-r^2/b_p^2) \; .
\end{eqnarray}
where $b_p(z)$ is the radius of the plume. 
We assume, as in RZ95, an isentropic and polytropic envelope structure,
hence
\begin{eqnarray}
\label{profil}
 \rho(z) &=& \rho_0 \, (z/z_0)^q \; , \nonumber \\
 T(z)    &=& T_0  \, (z/z_0) \; ,
\end{eqnarray}
where $q$ is the polytropic coefficient. $\rho_0$ and $T_0$ are the density and
temperature at depth $z=z_0$, and $z_0$ is the reference depth that corresponds
to the base of the convective region.

In Fig.~\ref{auto_sim}, we show that the mean velocity of upflow and downflow
in the quasi-adiabatic convection zone both obey a power law in $(z/z_0)^r$.
We therefore assume a power law for the mean velocity of the downflow (i.e.,
the plumes). Then
\begin{equation}
   <w>_d ~=~ w_{d0} \, \Big(\frac{z}{z_0}\Big)^r
\label{power_law}
\end{equation}
with
\begin{equation}
   w_{d0}=\left(\frac{12 F}{\beta_0^2 \pi \rho_0 g z_0^2}\right)^{1/3} \; ,
\end{equation}
(RZ95), where $r=(-q+1)/3$, $\beta_0=3\alpha/(q+2)$, and $\alpha=0.083$ is the
entrainment constant for a Gaussian profile \citep{Turner86}. $F$ is the
convective energy flux and $g$ is the gravitational acceleration. In
Table~\ref{table:1}, we list solar values of the previously introduced
parameters taken from RZ95. These values are used in the present paper except
for $F$, which is taken from the 3D numerical simulation (as explained below). 
For a monoatomic perfect gas, one has $q=3/2$, hence $r=-1/6$. However, our
3D numerical simulations indicate a value of $r$ closer to 0. The reason is likely
that there is radiative cooling. Hence, $\gamma > \Gamma=c_P/c_V$, where
$\gamma$ is the polytropic index ($q=1/(\gamma-1)$).

Following \citet{RZ95}, we assume that all the convective energy flux
is transported by the plume, thus
\begin{equation}
  F = L_{\odot}/N \; ,
\end{equation}
where $N$ is the number of plumes in the shell at $h=R_{\odot}-z$. We find
$N \approx 6.10^6$ from the 3D numerical simulation. 
To obtain such a result, one has to use the relation between $a$ and $N$:
\begin{equation}
\label{N}
  a = N\pi b_p^2 / 4\pi h^2 \; ,
\end{equation}
where ($a$) is mean fractional area of the upflow, $h=R_{\odot}-z$, and $b_p$ is the
radius of a plume. ($b_p$) and $a$ are taken from  the 3D  numerical simulation.
We assume $a=0.7$, as taken from Fig.~\ref{ab}, which shows that the mean
fractional area $a$ is roughly constant in the quasi-adiabatic convection zone.
The plume radius, $b_p$, is estimated at the top of the simulated box, which
corresponds to the photosphere.

\begin{table}
\caption{Solar values of plume model parameters (from RZ95)}
\label{table:1}
\centering
\begin{tabular}{c c}
\hline\hline
   $\beta_0$   & $\approx 0.1$  \\
   $\rho_0$    & $190$~kg~m$^{-3}$  \\
   $L_{\odot}$ & $3.9~10^{26}$~W  \\
   $z_0$       & $\approx 2.10^8$~m  \\
   $g_{\odot}$ & $270$~m~s$^{-2}$ \\
\hline
\end{tabular}
\end{table}

At this stage, we have modeled the downdrafts, but not yet the updrafts. The
3D numerical simulations show that mean velocities of upflow and downflow
obey the {\it same} power law (Fig.~\ref{auto_sim}). This can be explained as
follows: from the conservation of the mass flux one has
\begin{eqnarray}
\label{conserv_mass}
  <\rho w> &=& a <\rho w>_u + ~(1-a)< \rho w>_d \; =  \; 0 \; .
\end{eqnarray}
Fluctuating parts of densities in up and downflows are neglected such 
that $\rho_u \approx <\rho>_u$ and
$\rho_d \approx <\rho>_d$ (see Fig.~2b of RZ95). Thus,
\begin{equation}
  <w>_u ~=~ -\frac{(1-a)}{a} \frac{<\rho>_d}{<\rho>_u} <w>_d \; .
\end{equation}
Then, assuming that $<\rho>_u=\rho_{u0} (z/z_0)^q$ and $<\rho>_d=\rho_{d0}(z/z_0)^q$ obey
the same power law as in Eq.~(\ref{profil}):
\begin{eqnarray}
  <w>_u ~\approx~ -\frac{(1-a)}{a} \frac{\rho_{d0}}{\rho_{u0}} <w>_d \; .
\end{eqnarray}
$\rho_{u0}$ and $\rho_{d0}$ are the values at the reference depth $z_0$. 
We set $a \approx 0.7$ (see Fig.~\ref{ab}), which is the value obtained in the
quasi-adiabatic zone from the 3D numerical simulation.
Assuming further that $\rho_{d0}/\rho_{u0} \approx 1$, one obtains
\begin{equation}
  |<w>_u| \approx 0.45 ~|<w>_d| \; ,
\end{equation}
which is approximately what is seen in Fig.~\ref{auto_sim}.

\subsubsection{Skewness $S_w$ and the fourth-order moment}

\begin{figure}[t]
\begin{center}
\includegraphics[height=6cm,width=9cm]{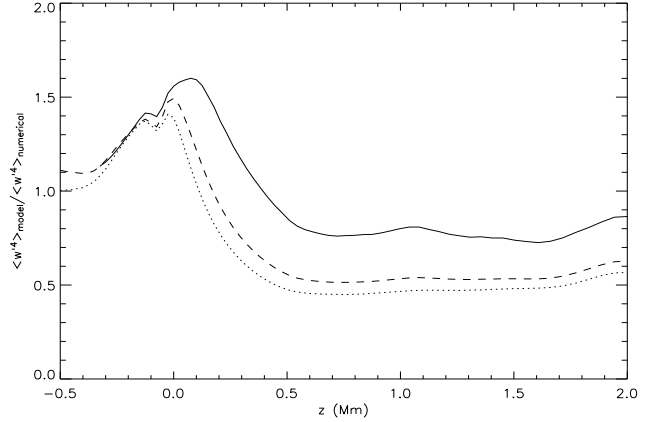}\\
\caption{Fourth-order moment $<w'^4>$ as a function of depth $z$ normalized to
    the FOM calculated directly from numerical simulations. The solid line
    shows $<w'^4>$ calculated using the CMP model, the dashed line is the
    moment as obtained from Eq.~(\ref{inter}) with Eq.~(\ref{S_2}) for $S_w$,
    and the dotted line is the QNA, Eq.~(\ref{w4_QNA}).}
\label{result_w4}
\end{center}
\end{figure}

We use Eq.~(\ref{Sw_eff}) for the skewness with
\begin{equation}
\label{delta_w}
  \delta w = (<w_u>-<w_d>) \approx 1.45 \, w_{d0} \; .
\end{equation}
The vertical depth of the computation box is narrow in comparison with the
reference depth $z_0$, thus $\delta w$ varies only weakly with $z$. Hence, we
assume $r=0$ in the solar case. 
The fourth-order moment $<w'^4>$ can then be computed by means of the interpolated
relation Eq.~(\ref{inter}). In Fig.~\ref{result_w4}, we show the resulting
$<w'^4>$. The CMP clearly is an improvement compared to the QNA and the TFM expression for $S_w$, Eq.~(\ref{S_2}) combined with Eq.~(\ref{inter}), 
by at least a factor two in the quasi-adiabatic zone. 
The FOM in the superadiabatic zone is overestimated. Indeed, as mentioned
above, the CMP is not able to describe such a zone mainly because the
assumptions of Sect.~\ref{turb} are not valid.
Note that it is possible to use the same procedure to compute any
other third- and fourth-order moment.

\subsubsection{Determination of $\delta \theta$}
\label{delta_theta}

Similarly to the procedure in the previous section, we evaluate
$<\tilde{\theta}'^2>$ with the help of Eqs.~(\ref{ordre_2_CMP}) and
(\ref{S_teta_CMP}). We therefore need to determine $\delta \theta$. The
temperature profile is more sensitive to departure from adiabaticity than the
velocity profile. It is therefore not suitable to assume an isentropic
 envelope. Such an approximation can still be used in the downflow, but not
for the upflow, which is far from being adiabatic due to radiative cooling. Then,
for the sake of simplicity, we assume a power law to obtain $\delta \theta$:
\begin{equation}
\label{power_law_theta}
  \delta \theta \approx \delta \theta_0 \Big(\frac{z}{z_0}\Big)^m \; .
\end{equation}
\begin{figure}[t]
\begin{center}
\includegraphics[height=6cm,width=9cm]{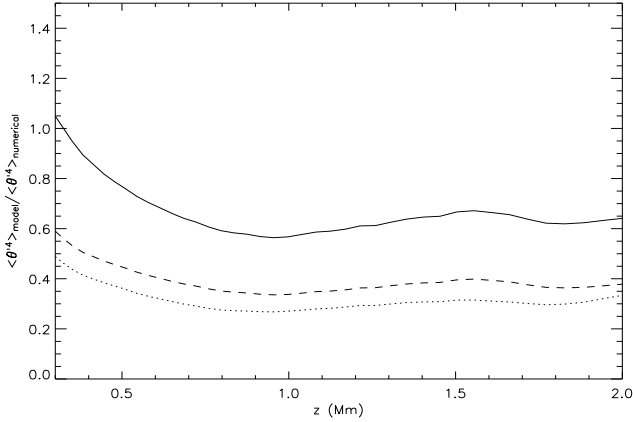}\\
\caption{Fourth-order moment $<\theta'^4>$ as a function of depth $z$
    normalized to the directly numerically calculated FOM. In solid lines
    the moment stems from $<\theta'^4>$ calculated using the CMP model,
    the dashed line is the moment as obtained from Eq.~(\ref{inter})
    and Eq.~(\ref{S_2}) for $S_{\theta}$ and the dotted line is the QNA.}
\label{result_teta4}
\end{center}
\end{figure}
For $z > 1$~Mm in the simulated box ($z=0$~Mm denotes the photosphere),
one derives $m=-1.5$, $\delta \theta_0 \approx 170$~K from the 3D numerical
simulation. Using the power law (Eq.~(\ref{power_law_theta})) with $m=-1.5$, the
skewness $S_{\theta}$ can be calculated using Eq.~(\ref{S_teta_CMP}). 
In Fig.~\ref{result_teta4}, we present the fourth-order moment $<\theta'^4>$
computed using the CMP, and as expected, the description of the FOM is improved. 
In the deeper part of the convection zone (i.e., the adiabatic region),
$\delta \theta$ is easier to model because Eq.~(\ref{profil}) can be used
and the difference $\delta \theta$ becomes a power law. 
From Eqs.~(\ref{no_inter_TOM}), (\ref{inter}), (\ref{Sw_eff}), and (\ref{S_teta_CMP}) all the
 third- and fourth-order moments can be modelled with the CMP.

\subsubsection{Summary: the CMP in a nutshell}

In practice, one uses the CMP  to compute $<w'^4>$ by means of the
interpolation formula Eq.~(\ref{inter}), where the second-order moment $<w'^2>$ is
supposed to be known and where the skewness $S_w$ is computed from
Eq.~(\ref{Sw_eff}). In the latter expression, $\delta w$ is determined using
the plume model through Eq.~(\ref{delta_w}) and using Eqs.~(\ref{power_law}) to 
(\ref{N}) with appropriate values of parameters for the case studied
(in the present paper we used the values from Table~\ref{table:1}, 
which are suitable for the solar case). 
 Here, $a(z)$, $N$, $b_p$, and other input quantities are taken from the
3D numerical simulation. 
When the CMP is used to obtain the other third- and fourth-order moments,
additional quantities have to be determined, namely $b$ and $m$ in
Eq.~(\ref{power_law_theta}) for $S_{\theta}$ (see Eq.~(\ref{S_teta_CMP})). 

\section{Conclusions}

 With the help of 3D numerical simulations of the upper part of
the solar convective region, we have shown that the QNA and the TFM 
fail to describe the fourth-order velocity and temperature correlation
moments, if merely used on their own. These results confirm KR2006 and
geophysical studies \citep{GH2002} and led us to generalize the TFM in order
to take the effects of the turbulent properties of the up- and downflows
explicitly into account (GTFM). We point out that the GTFM can be used in
other contexts than the solar one as long as the convective system can be
described with two turbulent flows.

 One might wonder whether it is likely that the CMP and the model for $p$~mode
excitation developed in Paper~II are generally applicable to solar-like stars.
To answer this question requires further work, but results on important
ingredients of these models are encouraging. The case of convection in the
planetary boundary layer of the atmosphere of the earth was already discussed in
GH2002. Their interpolation model for FOMs has meanwhile been investigated for
the case of convection in the ocean \citep{Losch04} and solar granulation
(\citealt{KR2006}, who also study the case of a K~dwarf; preliminary results
were published in \citealt{Kupka05}). We corroborate the latter here with
simulations for solar granulation based on more realistic boundary conditions.
The overall conclusion that can be drawn from these studies is that, at least
away from the boundary layers of convection zones, the FOMs in purely convective
flows can be estimated according to the interpolation model by GH2002 with an
accuracy typically in the range of 20\% to 30\%, whereas the QNA is off by
a factor of two to three. For the superadiabatic layer, the discrepancies of the QNA 
remain the same in any case of the same size.

We focused here on the solar case, more precisely a region that is nearly adiabatic, 
just below the superadiabatic
zone where the acoustic modes are excited.
As indicated by the 3D simulations, the coherent downdrafts,
called plumes, are more turbulent than the upflow. In addition, we use the
plume model developed by RZ95 to estimate the upward and downward mean
velocities. With these additional approximations, the GTFM yields a closure
model, the CMP, which can be applied in the quasi-adiabatic zone (located just
below the superadiabatic one). Comparisons of calculations based on the CMP
 with direct calculations from the 3D numerical simulations show a good
agreement. Hence, the CMP provides an analytical closure for third- and
fourth-order moments. These moments are expressed in a simple way and
require only the knowledge of the second-order moments and the parameters
of the plume model. We stress that the CMP involves four parameters: the
number of plumes in the considered shell (i.e., near the photosphere), the
exponent of the power law for the mean vertical velocity of plumes, the law to
describe the temperature difference between the two flows, and the mean
fractional area of the updrafts and hot drafts.

A study of the dependence of the results on these parameters is in progress. 
For instance, an increase of $a$ will imply an increase of $S_w$ in Eq.~(\ref{Sw_eff}), 
 and hence of the fourth-order moment $<w'^4>$. Nevertheless,
it is extremely difficult to deduce the behaviour of the system, since from
Eq.~(\ref{conserv_mass}) a variation of $a$ changes the velocities of the flows. 
Instead, one could use a set of numerical simulations to study the effect
of a change of the parameter $a$. 
In a companion paper, we use the CMP in a semi-analytical approach to
calculate the power supplied to the solar $p$~modes. It is found that the
power is quite significantly affected by the adopted closure model.

Our final aim is to apply the CMP to the study of stochastic excitation of
solar-like $p$~modes in stars other than the Sun. It will be necessary to assess
the validity of the CMP approximations to extend their application to
stellar conditions different from the solar case. This will also
require investigating the dependence of the parameters entering the CMP, for
instance, on the effective temperature of the star (work which is in progress). 
As pointed out in Sect.~\ref{CMP}, the CMP is valid only in the
quasi-adiabatic zone due to the power laws used to model the plume dynamics.
This will be discussed further in the companion paper in which the present
model will be used in the superadiabatic zone in order to propose a new
closure for the calculation of stellar $p$~modes.

Finally, 
we note that in the present work we do not take the effect of differential 
rotation and meridional circulation into account. However, recent helioseismic investigations \citep{Schou02,Zhao04} have shown that 
variability of those large-scale flows gradually affects  
wavelength and frequencies, leading to a redistribution of the observed power spectrum \citep{Sher05,Hindman05}.  
Hence, it could have an indirect effect on the amplitudes of $p$~modes. 
Furthermore, large-scale laminar non-uniform flows  can have a significant effect on the 
formation of the coherent structures and intrinsic turbulence \citep{Miesch00,Brun02,Rempel05}. 
To what extent they can affect solar $p$~mode amplitudes,   
through the closure model and the Reynolds stresses, remains to be investigated.\\ 
 

\begin{acknowledgements}

We are indebted to J.~Leibacher for his careful reading of the manuscript and his helpful remarks.
We thank J.P~Zahn and F.~Ligni\`eres for useful discussions and comments.
FK is grateful to V.M.~Gryanik and J.~Hartmann for discussions on their model and
their observational data. We also thank the anonymous referee for valuable comments that 
helped to improve the manuscript.

We thank {\AA}. Nordlund and R.~F. Stein for making their code 
available to us.  Their
code was made at the National Center for Supercomputer 
Applications and
Michigan State University and supported by grants from NASA and NSF.\\

\end{acknowledgements}

\newpage
\appendix

\section{Cross-correlation moments} \label{Sect_app_A}

As explained in Sect.~\ref{Sect_GTFM_formulation}, we provide the cross-correlation moments:
\begin{eqnarray}
<w' \theta'> &=&  a_{uh} <\tilde{w}'\tilde{\theta}'>_{u,h}+~a_{uc}<\tilde{w}'\tilde{\theta}'>_{u,c} \nonumber \\	  
             &+& a_{dc}<\tilde{w}'\tilde{\theta}'>_{d,c}+~ a_{dh}<\tilde{w}'\tilde{\theta}'>_{d,h} + ~\eta \, \delta w \delta \theta  \\  \nonumber \\
<w'^2 \theta'> &=& a_{uh}<\tilde{w}'^2 \tilde{\theta}'>_{u,h}+~a_{uc}<\tilde{w}'^2\tilde{\theta}'>_{u,c}  			               \nonumber \\        &+& a_{dh}<\tilde{w}'^2\tilde{\theta}'>_{d,h}
                   + ~a_{dc}<\tilde{w}'^2\tilde{\theta}'>_{d,c}  \nonumber \\
		    &+& \beta_1 \, \delta \theta  + \beta_2 \, \delta w + \beta_3 \, \delta w^2 \delta \theta   \\ \nonumber \\
<w' \theta'^2>   &=& a_{uh}<\tilde{w}'\tilde{\theta}'^2>_{u,h}+~a_{uc}<\tilde{w}'\tilde{\theta}'^2>_{u,c}
	\nonumber \\	      &+& a_{dh}<\tilde{w}'\tilde{\theta}'^2>_{d,h}
		      +~ a_{dc}<\tilde{w}'\tilde{\theta}'^2>_{d,c} \nonumber \\ &+& \gamma_1 \, \delta w + \gamma_2 \, \delta \theta + \gamma_3 \, \delta w \delta \theta^2  \\
    \nonumber \\
<w'^2 \theta'^2> &=& a_{uh}<\tilde{w}'^2\tilde{\theta}'^2>_{u,h}+~a_{uc}<\tilde{w}'^2\tilde{\theta}'^2>_{u,c} \nonumber \\	     
                 &+& a_{dh}<\tilde{w}'^2\tilde{\theta}'^2>_{d,h}+~ a_{dc}<\tilde{w}'^2\tilde{\theta}'^2>_{d,c} \nonumber \\ 
                 &+& \phi_1 \, \delta \theta + \phi_2 \, \delta w + \phi_3 \, \delta \theta^2 + \phi_4 \, \delta  w^2 \nonumber \\
		 &+& \phi_5 \, \delta w \delta \theta + \phi_6 \, \delta w^2 \delta \theta^2
\end{eqnarray}
where:
\begin{eqnarray}
\eta &=& \Big[a_{uh}(1-a)(1-b)-a_{uc}b(1-a)+a_{dc}ab-a_{dh}a(1-b) \Big] \nonumber \\
\beta_1 &=& \Big[a_{uh}(1-b)-a_{uc}b\Big] \, <\tilde{w}'^2>_u  \nonumber \\
        &+& \Big[a_{dh}(1-b)-a_{dc}b\Big] \, <\tilde{w}'^2>_d  \nonumber \\
\beta_2 &=& 2 \Big[a_{uh}<\tilde{w}' \tilde{\theta}'>_{u,h}+a_{uc}<\tilde{w}' \tilde{\theta}'>_{u,c}-a(<w'\theta'>-\eta)\Big] \nonumber \\
\beta_3 &=& \Big[a_{uh} (1-a)^2(1-b) -a_{uc} b(1-a)^2+a_{dh}a^2(1-b)-a_{dc}a^2b\Big] \nonumber \\
\gamma_1 &=& \Big[a_{uh}(1-a)-a_{dh}a\Big] \, <\tilde{\theta}'^2>_{h}  \nonumber \\
        &+& \Big[a_{uc}(1-a)-a_{dc}a\Big] \, <\tilde{\theta}'^2>_{c}  \nonumber \\
\gamma_2 &=& 2 \Big[a_{uh}<\tilde{w}' \tilde{\theta}'>_{u,h}+~a_{dh}<\tilde{w}' \tilde{\theta}'>_{d,h}-~b(<w'\theta'>-\eta)\Big] \nonumber \\
\gamma_3 &=& \Big[a_{uh} (1-a)(1-b)^2 -a_{uc} b^2(1-a) - a_{dh}a(1-b)^2-a_{dc}ab^2\Big] \nonumber \\
\phi_1 &=& 2 \Big[a_{uh} <\tilde{w}'^2 \tilde{\theta}'>_{u,h} +~a_{dh} <\tilde{w}'^2 \tilde{\theta}'>_{d,h} \nonumber \\          
&-& b(<w'^2 \theta'> - ~\beta_1 -~\beta_2 -~\beta_3) \Big] \nonumber \\
\phi_2 &=& 2 \Big[a_{uh} <\tilde{w}' \tilde{\theta}'^2>_{u,h} +~a_{uc} <\tilde{w}'^2 \tilde{\theta}'>_{u,c} \nonumber \\        
&-& a(<w' \theta'^2> -~ \gamma_1 -~\gamma_2 -~\gamma_3) \Big] \nonumber \\
\phi_3 &=& \Big[a_{uh}(1-b)^2-a_{uc}b^2\Big] \, <\tilde{w}'^2>_{u}  \nonumber \\
        &+& \Big[a_{dh}(1-b)^2-a_{dc}b^2\Big] \, <\tilde{w}'^2>_{d} \nonumber \\
\phi_4 &=& \Big[a_{uh}(1-a)^2-a_{dh}a^2\Big] \, <\tilde{\theta}'^2>_{h}  \nonumber \\
        &+& \Big[a_{uc}(1-a)^2-a_{dc}a^2\Big] \, <\tilde{\theta}'^2>_{c}  \nonumber \\
\phi_5 &=& 4 \Big[a_{uh}(1-a)(1-b)<\tilde{w}'\tilde{\theta}'>_{u,h} -~ a_{uc}b(1-a) <\tilde{w}'\tilde{\theta}'>_{u,c} \nonumber \\ &-& a_{dh} a(1-b) <\tilde{w}'\tilde{\theta}'>_{d,h} +~ a_{dc} ab <\tilde{w}'\tilde{\theta}'>_{d,c} \Big] \nonumber \\
\phi_6 &=& \Big[a_{uh} (1-a)^2(1-b)^2 +a_{uc} (1-a)^2 b^2 + a_{dh} a^2(1-b)^2 \nonumber \\ &+& a_{dc} a^2 b^2 \Big] \nonumber \; .
\end{eqnarray}

\newpage


\end{document}